\documentclass[preprintnumbers,amsmath,amssymb,nofootinbib]{revtex4}

\usepackage{graphicx}
\usepackage{epsfig}
\usepackage{color}

\newcommand{\beq}{\begin{equation}}
\newcommand{\eeq}{\end{equation}}
\newcommand{\bea}{\begin{eqnarray}}
\newcommand{\eea}{\end{eqnarray}}
\newcommand{\bwd}{\begin{widetext}}
\newcommand{\ewd}{\end{widetext}}

\begin{document}

\title{Differentiable self-consistent space-charge simulation for accelerator design}


\author{Ji Qiang}
\email{jqiang@lbl.gov}
\affiliation{Lawrence Berkeley National Laboratory, Berkeley, CA 94720, USA}

\begin{abstract}
The nonlinear space-charge effects in a high intensity or high brightness accelerator
can have a significant impact on the beam properties through the accelerator.
These effects are included in the accelerator design via
self-consistent multi-particle tracking simulations. 
In order to study the sensitivity of the final beam's properties 
with respect to the accelerator design parameters, one has to carry out the
time-consuming space-charge simulation multiple times. 
In this paper, we propose a differentiable self-consistent space-charge
simulation model that enables the study of the beam sensitivity through 
only one simulation.
Such a model can also be used with gradient-based numerical optimizers for
accelerator design optimizations including the self-consistent space-charge effects.
\end{abstract}

\maketitle

\section{Introduction}

The nonlinear space-charge effects from the Coulomb interaction inside 
a charged particle beam can have a significant impact on the beam 
transport
through an accelerator by causing beam emittance growth, halo formation, and even 
particle losses. These effects are normally studied in the accelerator design 
via self-consistent simulations.
To simulate the space-charge effects self-consistently, 
multi-particle tracking with a particle-in-cell method
has been widely employed
in the accelerator community~\cite{friedman,takeda,machida,jones,impact,track,galambos,franchetti03,impact-t,amundson,opal}.
However, none of these codes directly calculates the derivatives of
the beam property with respect to the accelerator machine parameters.

The derivatives of the beam property with respect to the accelerator
machine parameters are important in the accelerator design.
These derivatives provide quantitative measurements of the sensitivity
of the final beam property with respect to the machine parameters.
Such sensitivities can be used to set the tolerance limits of the
corresponding machine parameters.
In a typical machine design, the sensitivity can be obtained by running the
space-charge simulation multiple times, each time with a small change of a
single machine parameter. A numerical scheme such as the finite difference
method is used to calculate the derivative of the beam property
with respect to that parameter. In an accelerator, there can be
hundreds and thousands machine parameters. To compute 
the derivatives with respect to all machine parameters 
using the finite difference method will involve hundreds
and thousands self-consistent space-charge simulations and can be
very time consuming. In this paper, we propose a differentiable
self-consistent space-charge simulation model. The derivatives
of the final beam property with respect to the entire machine parameters
can be obtained through a single simulation. Such a differentiable
space-charge simulation can also be integrated into a gradient-based
optimizer for accelerator design optimization.

The differentiable simulator is a simulator that can automatically
compute derivatives of the simulation result with respect to its
input parameters.  
This can be done through automatic differentiation~\cite{ad} that
has been widely used in the artificial intelligent/machine learning (AI/ML)
community to train a neural network through gradient-based optimization method.
A number of AI/ML frameworks such as PyTorch~\cite{pytorch} and 
TensorFlow~\cite{tensorflow} include
this capability.
Recently, a differentiable model was developed to optimize magnet design
by implementing the non-parametric version of the Preisach model output 
using the PyTorch framework~\cite{roussel}. 
In this paper, instead of using the automatic differentiation framework
of the AI/ML community, we proposed a differentiable
self-consistent space-charge simulation model using the truncated
power series algebra that was
developed in the accelerator community.
This method can also be implemented in many other available self-consistent 
space-charge
codes for high intensity, high brightness accelerator design study.

The truncated power series algebra (TPSA)
was first introduced by Berz in 1989 for accelerator applications~\cite{berz0}.
Since then, it has been used to calculate
transfer maps of beam line elements in several optics codes~\cite{yan,berz,forest,cosy}.
It was also used to calculate the space-charge potential/fields in
the fast multipole method~\cite{zhang2011,nissenthesis,zhangthesis} 
and to solve the Poisson's
equation~\cite{erdelyi2015}.
Recently, it was also used to extract transfer maps in 
the presence of space-charge effects~\cite{gee2014,nissen,qiang2019}.
However, so far, these applications of the TPSA have no direct connections
to the control parameters in the accelerator design.
In this study, we present a new application of the TPSA to the accelerator
design in the self-consistent space-charge simulation.
In this differentiable space-charge model, the final beam property
such as emittance is connected to the accelerator 
control parameters. This enables to study
the sensitivity of the final beam property with respect to these
control parameters through only one simulation, which
is different from the other uncertainty quantification methods such as
the surrogate model method in which a number of
simulations have to be used to train the model~\cite{adelmann2019}.

The organization of this paper is as follows: after the introduction, we introduce the
truncated power series algebra in Section II,
present
a differentiable space-charge model in Section III, 
two application examples in 
Section IV, and draw conclusions in Section VI.

\section{Truncated power series algebra}

The truncated power series algebra (TPSA) is an effective tool to calculate
derivatives of a function with respect to its variables using
an algebraic method. 
Consider the Taylor series approximation of a one-dimensional function $f(x)$
at a point $x_0$,
\begin{eqnarray}
f(x) & = & f(x_0) + (x-x_0)f'(x_0)+\frac{1}{2!}(x-x_0)^2f''(x_0)+\cdots+
\frac{1}{N!}(x-x_0)^Nf^(N)(x_0)
\end{eqnarray}
the derivatives in the above equation can be calculated using
a numerical finite difference method, for example,
\begin{eqnarray}
	f'(x_0) & = & \frac{f(x_0+\delta)-f(x_0)}{\delta} + O(\delta)
\end{eqnarray}
Such a way to calculate the derivative introduces numerical errors
and requires multiple function evaluations for a multi-variable function.
In the truncated power series algebra method, the derivative up to
$N^{th}$ order can be regarded as a point in a function space spanned by
the bases:
\begin{eqnarray}
{1,(x-x_0),\frac{1}{2!}(x-x_0)^2,\cdots,\frac{1}{N!}(x-x_0)^N}
\end{eqnarray}
These derivatives can be represented as a vector:
\begin{eqnarray}
	Df_{x_0} & = & [f(x_0),f'(x_0),f''(x_0),\cdots, f^{(N)}(x_0)]
\end{eqnarray}
Such a vector is also called a TPSA variable.
For example, for a constant $c$, its derivative representation
is $Dc = [c,0,0,\cdots,0]$, and for a variable $x$, $Dx=[x,1,0,\cdots,0]$.
By using the above derivative vector representation,
the derivatives of function with respect to its variable can
be written as the function of that derivative vector, i.e.
\begin{eqnarray}
Df_{x} & = & f(Dx)
\end{eqnarray}
The computing of the derivatives of a function becomes the algebraic
function evaluations.

The evaluation of the derivative vector inside a function can
be broken down as the operations of addition and multiplication.
Given two derivative vectors 
$Df_{x_0}=[f(x_0),f'(x_0),f''(x_0),\cdots, f^{(N)}(x_0)]
=[a_0,a_1,a_2,\cdots,a_N]$
and $Df_{x_1}=[f(x_1),f'(x_1),f''(x_1),\cdots, f^(N)(x_1)]
=[b_0,b_1,b_2,\cdots,b_N]$, the sum of two vectors will be:
\begin{eqnarray}
Df_{x_0} +Df_{x_1}& = & [a_0+b_0,a_1+b_1,a_2+b_2,\cdots,a_N+b_N]
\end{eqnarray}
The multiplication of these two vectors will be,
\begin{eqnarray}
Df_{x_0}\times Df_{x_1}& = & [a_0b_0,a_0b_1+a_1b_0,a_0b_2+2a_1b_1+a_2b_0,\cdots,c_N]
\end{eqnarray}
where $c_N = \sum_{k=0}^N \frac{N!}{k!(N-k)!}a_kb_{N-k}$.
Using the above rules of addition and multiplication, the operation
of a derivative vector inside a function can be calculated algebraically. 
For example, the reciprocal of a derivative vector $1/Df_{x_0}$ can
be calculated as:
\begin{eqnarray}
Df_{x_0}^{-1}& = & [a_0,a_1,a_2,\cdots,a_N]^{-1} \nonumber \\
	& = & [\frac{1}{a_0},-\frac{a_1}{a_0^2},\frac{2a_1^2}{a_0^3}-\frac{a_2}{a_0^2},\cdots]
\end{eqnarray}
For a concrete example, $f(x) = \frac{1}{1+x+x^2}$,
one can use the above derivative vector operations to 
obtain the first and the second derivative of this
function at $x=1$. 
That is, $x=1$, $Dx = [1,1,0]$, and 
\begin{eqnarray}
Df_1 & = &f(D1) \nonumber \\
	& = & \frac{1}{1+[1,1,0]+[1,1,0]^2} \nonumber \\
	& = & [\frac{1}{3},-\frac{1}{3},\frac{4}{9}]
\end{eqnarray}
This yields the first derivative $f'(1) = -\frac{1}{3}$ and
the second derivative $f''(1) = \frac{4}{9}$. The truncated
power series algebra changes the calculation of 
the derivatives of a function 
with respect to its individual variable into
the evaluation of a function of derivative vector, i.e. a function of TPSA variable.

The above single variable function example can be extended
to a multiple variable function with the more complicated multiplication
rule for the derivative vector~\cite{chao}. The truncated power series algebra
libraries have been developed to handle some general 
special functions such as the exponential function and 
the trigonometry function~\cite{hao,he}.

\section{Differentiable self-consistent space-charge model}

The above truncated power series algebra is used to develop
a differentiable self-consistent space-charge model using TPSA variables.
The self-consistent space-charge simulation is done by solving
the following Hamilton equations:
\begin{eqnarray}
	\frac{d {\bf r}_i}{d s} & = & \frac{\partial H}{\partial {\bf p}_i} \\
	\frac{d {\bf p}_i}{d s} & = & -\frac{\partial H}{\partial {\bf r}_i} 
\end{eqnarray}
where $H({\bf r}_1,{\bf p}_1,{\bf r}_2,{\bf p}_2,\cdots,s)$ denotes the Hamiltonian of the system, 
and ${\bf r}_i$ and ${\bf p}_i$ denote canonical
coordinates and momenta of particle $i$, respectively. 
Let $\zeta$ denote a 6N or 4N-vector of coordinates,
the above Hamilton's equation can be rewritten as:
\begin{eqnarray}
	\frac{d \zeta}{d s} & = & -[H, \zeta] 
\end{eqnarray}
where [\ ,\ ] is the Poisson bracket. A formal solution for the above equation
after a single step $\tau$ can be written as:
\begin{eqnarray}
	\zeta (\tau) & = & \exp(-\tau(:H:)) \zeta(0)
\end{eqnarray}
Here, we have defined a differential operator $:H:$ as $: H : g = [H, \ g]$, 
for arbitrary function $g$. 

For a coasting beam, the Hamiltonian can be written as $H = H_1+H_2$, where
\begin{eqnarray}
	H_1 & = & \sum_{i=1}^{N_p} {\bf p}_i^2/2 + \sum_{i=1}^{N_p} q A_z({\bf r}_i)
\end{eqnarray}
where $A_z$ denotes the
longitudinal vector potential associated with the external focusing fields and 
\begin{eqnarray}
	H_2 & = & \frac{K}{4} \sum_{i=1}^{N_p} \sum_{j=1}^{N_p} \varphi({\bf r}_i,{\bf r}_j)
	\label{htot2}
\end{eqnarray}
where $K = q I/(2\pi \epsilon_0 p_0 v_0^2 \gamma_0^2)$ is the generalized 
perveance,
$q$ is the charge of particle, $I$ is the beam current, $\epsilon_0$ is the dielectric 
constant in vacuum, $p_0$ is the momentum of the reference
particle, $v_0$ is the speed of the reference particle, $\gamma_0$ is
the relativistic factor of the reference particle, and $\varphi$ is the 
space charge Coulomb interaction potential.
In this Hamiltonian, the effects of the direct electric potential and the
longitudinal vector potential are combined together.
For a Hamiltonian that can be written as a sum of two terms $H =  H_1 + H_2$, an approximate
solution to the above formal solution can be written as:
\begin{eqnarray}
	\zeta (\tau) & = & \exp(-\tau(:H_1:+:H_2:)) \zeta(0) \nonumber \\
  & = & \exp(-\frac{1}{2}\tau :H_1:)\exp(-\tau:H_2:) \exp(-\frac{1}{2}\tau:H_1:) \zeta(0) + O(\tau^3)
\end{eqnarray}
Let $\exp(-\frac{1}{2}\tau :H_1:)$ define a transfer map ${\mathcal M}_1$ and
$\exp(-\tau:H_2:)$ a transfer map ${\mathcal M}_2$, 
for a single step, the above splitting results in a second order numerical integrator
for the original Hamilton's equation as:
\begin{eqnarray}
	\zeta (\tau) & = & {\mathcal M}(\tau) \zeta(0) \nonumber \\
    & = & {\mathcal M}_1(\tau/2) {\mathcal M}_2(\tau) {\mathcal M}_1(\tau/2) \zeta(0)
	+ O(\tau^3)
	\label{map}
\end{eqnarray}
For the external focusing with quadrupole magnets, the 
single step transfer map ${\mathcal M}_1$ in the focusing plane can be written as:
\begin{equation}
	{\mathcal M}_1(\tau)   =   \left( \begin{array}{cc}
		\cos(\sqrt{k}\tau) & \frac{1}{\sqrt{k}}\sin(\sqrt{k}\tau) \\
		-\sqrt{k}\sin(\sqrt{k}\tau) & \cos(\sqrt{k}\tau)
	\end{array} \right)
\end{equation}
and in the defocusing plane as:
\begin{equation}
	{\mathcal M}_1(\tau)   =   \left( \begin{array}{cc}
		\cosh(\sqrt{k}\tau) & \frac{1}{\sqrt{k}}\sinh(\sqrt{k}\tau) \\
		-\sqrt{k}\sinh(\sqrt{k}\tau) & \cosh(\sqrt{k}\tau)
	\end{array} \right)
\end{equation}
where $k$ is the normalized focusing strength $k=q g/p_0$ and $g$ is the 
magnetic field gradient.
For the space-charge Hamiltonian $H_2({\bf r})$, the single
step transfer map ${\mathcal M}_2$ can be written as:
\begin{eqnarray}
	{\bf r}_i(\tau) & = & {\bf r}_i(0) \\
	{\bf p}_i(\tau) & = & {\bf p}_i(0) - \frac{\partial H_2({\bf r})}{\partial {\bf r}_i} \tau
	\label{map2}
\end{eqnarray}

The electric Coulomb potential in the Hamiltonian $H_2$ can be obtained 
from the solution of the Poisson equation.
In the following, we assume that the coasting beam is inside a rectangular perfectly conducting pipe.
In this case, the two-dimensional Poisson's equation can be written as:
\begin{equation}
\frac{\partial^2 \phi}{\partial x^2} +
\frac{\partial^2 \phi}{\partial y^2} = - 4 \pi \rho
\label{poi2d}
\end{equation}
where
$\phi$ is the electric potential, and $\rho$ is the particle
density distribution of the beam.

The boundary conditions for the electric potential inside the rectangular 
perfectly conducting pipe are:
\begin{eqnarray}
	\label{bc1}
\phi(x=0,y) & = & 0  \\
\phi(x=a,y) & = & 0  \\
\phi(x,y=0) & = & 0  \\
\phi(x,y=b) & = & 0  
	\label{boundary}
\end{eqnarray}
where $a$ is the horizontal width of the pipe and $b$ is the vertical width
of the pipe. 

Given the boundary conditions in Eqs.~\ref{bc1}-\ref{boundary}, using
a Galerkin spectral approximation method, one obtains 
the space-charge Hamiltonian $H_2$ as~\cite{qiang2018}:
\begin{eqnarray}
	H_2  = 4\pi \frac{K}{ab} \frac{1}{N_p} \sum_{i=1}^{N_p} \sum_{j=1}^{N_p} \sum_{l=1}^{N_l} \sum_{m=1}^{N_m} 
	\frac{1}{\gamma_{lm}^2} \sin(\alpha_l x_j) \nonumber \\
	\sin(\beta_m y_j) \sin(\alpha_l x_i) \sin(\beta_m y_i)
\end{eqnarray}
The resultant one-step symplectic transfer map ${\mathcal M}_2$ of the 
particle $i$ with this Hamiltonian is given as:
\begin{eqnarray}
	p_{xi}(\tau) & = & p_{xi}(0) -
	\tau \frac{K}{2}
	\sum_{l=1}^{N_l} \sum_{m=1}^{N_m} \phi^{lm} 
	\alpha_l \cos(\alpha_l x_i) \sin(\beta_m y_i)  \nonumber \\
	p_{yi}(\tau) & = & p_{yi}(0) -
	\tau \frac{K}{2} 
	\sum_{l=1}^{N_l} \sum_{m=1}^{N_m} \phi^{lm}
	\beta_m \sin(\alpha_l x_i) \cos(\beta_m y_i)
\end{eqnarray}
where the space-charge potential in the spectral domain is given as:
\begin{eqnarray}
	\phi^{lm} & = & 4 \pi \frac{4}{ab} \frac{1}{N_p} \sum_{j=1}^{N_p} 
		\frac{1}{\gamma_{lm}^2} \sin(\alpha_l x_j) \sin(\beta_m y_j)
\end{eqnarray}
Here, both $p_{xi}$ and $p_{yi}$ are normalized by the reference particle momentum $p_0$.

In the differentiable space-charge simulation, the above 
space-charge model will be rewritten using TPSA variables.
The phase space coordinates 
${\bf r}_i$ and ${\bf p}_i$
of the particle $i$ will be replaced by the corresponding TPSA variables 
$D{\bf r}_i$ and $D{\bf p}_i$
defined in
the last section. The potential $\phi^{lm}$ is replaced by $D\phi^{lm}$
of the TPSA variable. The momentum updates after a single step due
to the space-charge effects are given by:
\begin{eqnarray}
	Dp_{xi}(\tau) & = & Dp_{xi}(0) -
	D\tau \frac{K}{2}
	\sum_{l=1}^{N_l} \sum_{m=1}^{N_m} D\phi^{lm} 
	\alpha_l \cos(\alpha_l Dx_i) \sin(\beta_m Dy_i)  \nonumber \\
	Dp_{yi}(\tau) & = & Dp_{yi}(0) -
	D\tau \frac{K}{2} 
	\sum_{l=1}^{N_l} \sum_{m=1}^{N_m} D\phi^{lm}
	\beta_m \sin(\alpha_l Dx_i) \cos(\beta_m Dy_i)
\end{eqnarray}
where the space-charge potential in TPSA variable is: 
\begin{eqnarray}
	D\phi^{lm} & = & 4 \pi \frac{4}{ab} \frac{1}{N_p} \sum_{j=1}^{N_p} 
		\frac{1}{\gamma_{lm}^2} \sin(\alpha_l Dx_j) \sin(\beta_m Dy_j)
\end{eqnarray}
The map corresponding to the external quadrupole field can also
be written in TPSA variable as:
\begin{equation}
	{\mathcal M}_1(\tau)   =   \left( \begin{array}{cc}
		\cos(\sqrt{Dk}D\tau) & \frac{1}{\sqrt{Dk}}\sin(\sqrt{Dk}D\tau) \\
		-\sqrt{Dk}\sin(\sqrt{Dk}D\tau) & \cos(\sqrt{Dk}D\tau)
	\end{array} \right)
\end{equation}
where $Dk$ is the TPSA variable of the quadrupole focusing strength,
and $D\tau$ is the TPSA variable of the step size that is
the quadrupole length divided by the number of steps. A similar expression
can be written in the defocusing plane of the quadrupole.

The charged particle beam initial distribution parameters,
beam energy, and current can also be 
written using TPSA variables if needed.
The final beam properties such as emittances are defined
using the TPSA variables. 
The horizontal emittance $D\epsilon_x$ is given as
\begin{eqnarray}
	D\epsilon_x & = & \sqrt{D<x^2>D<p_x^2>-(D<xp_x>)^2}
\end{eqnarray}
where
\begin{eqnarray}
	D<x^2> & = & \sum_{i=1}^{N_p} (Dx_i)^2 \\
	D<{p_x}^2> & = & \sum_{i=1}^{N_p} (Dp_{xi})^2 \\
D<{xp_x}> & = & \sum_{i=1}^{N_p} Dx_i Dp_{xi} 
\end{eqnarray}
The vertical emittance has a similar expression with $x$ replaced by $y$.

\begin{figure}[!htb]
   \centering
   \includegraphics*[angle=0,width=174pt]{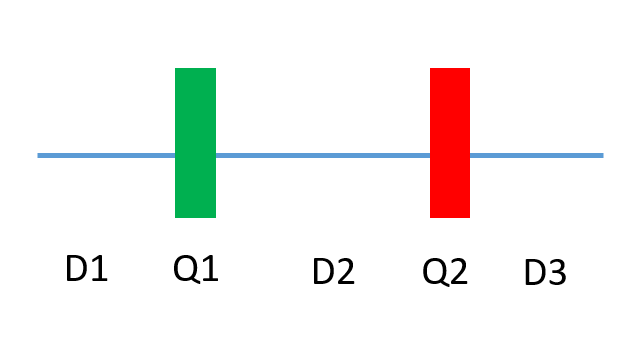}
   \caption{Schematic plot of a FODO lattice in the first application example.}
   \label{fig1}
\end{figure}
\begin{figure}[!htb]
   \centering
   \includegraphics*[angle=0,width=80mm,height=57mm]{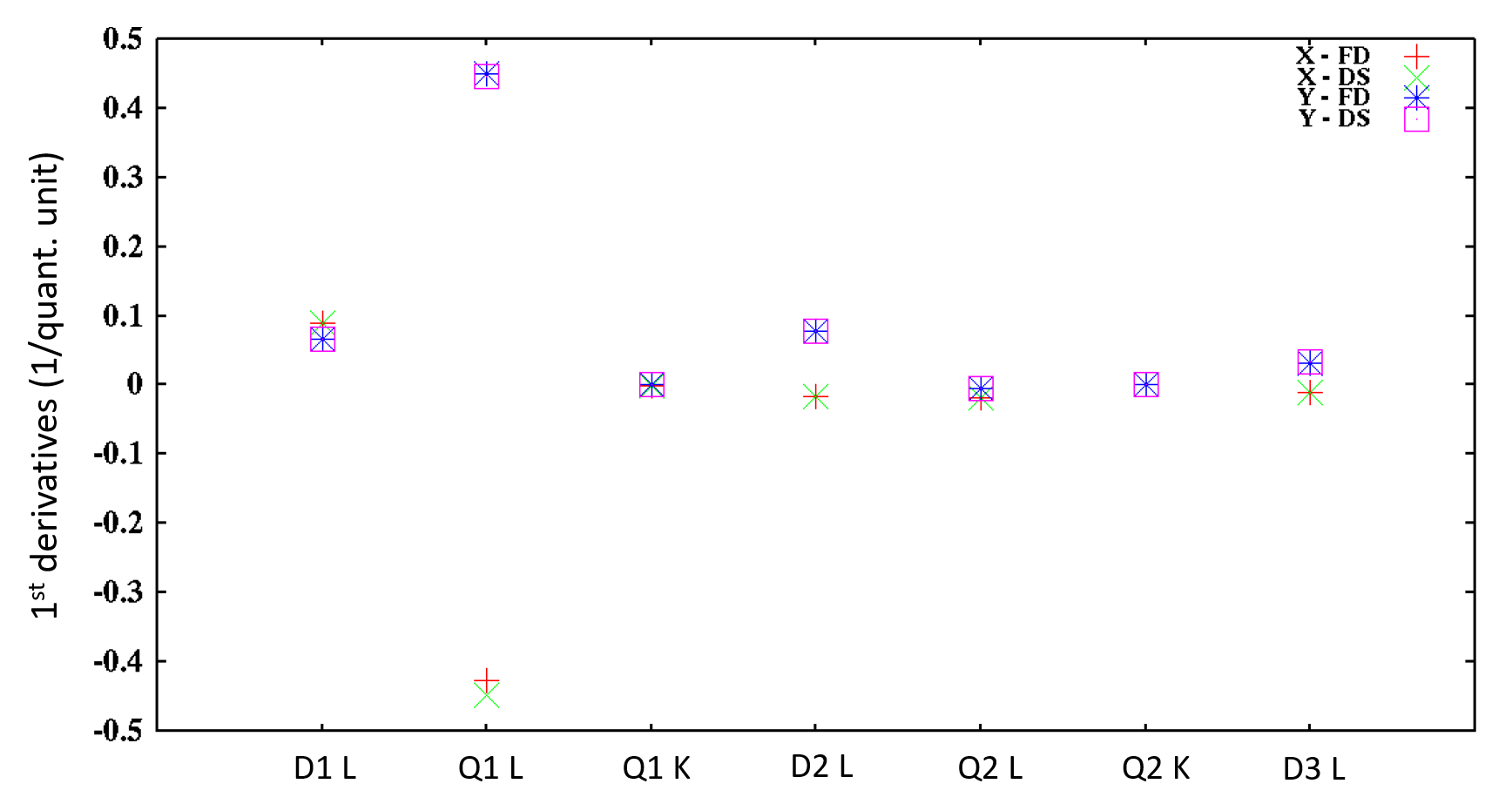}
   \caption{
Derivatives of final horizontal and vertical emittances
with respect to seven lattice parameters from the single
differentiable self-consistent space-charge simulation and from
the finite difference approximation to the first derivative.}
   \label{fig2}
\end{figure}

\section{Application examples}

In the following, we will use two application examples to illustrate
the above differentiable self-consistent space-charge simulation model:
one for parameter sensitivity study, and the other one for
design optimization study.

In the first illustrative application,
we studied sensitivity of the final emittances of a $1$ GeV coasting 
proton beam transporting through a transverse focusing FODO lattice 
inside a rectangular perfectly conducting pipe
with respect to the lattice parameters and the initial beam parameters.
A schematic plot of the FODO lattice is shown in Fig.~1.
It consists of a drift (D1) of $0.2$ meters, a quadrupole (Q1) of $0.1$ meters
and focusing strength $29.6/m^2$ for transverse focusing, 
another drift (D2) of $0.4$ meters, another quadrupole (Q2) of the same
length as the first quadrupole but opposite sign of focusing, and another drift (D3) of length $0.2$ meters. The rectangular pipe has an aperture size
of $13$ by $13$ millimeters. The zero current phase advance of the FODO lattice
is $87.0$ degrees.
The current of the proton beam is $200$ Amperes with $1$ mm mrad 
normalized emittance, which results in a depressed
phase advance of $63.1$ degrees.
The initial distribution is assumed to be a four-dimensional Gaussian
distribution given by:
\begin{eqnarray}
	f(x,p_x,y,p_y) & \propto & \exp{(-\frac{1}{2}( \frac{x^2}{\sigma_x^2} + 2xp_x\frac{\mu_{xp_x}}{\sigma_x\sigma_{p_x}}+\frac{p_x^2}{\sigma_{p_x}^2}))} 
	\exp{(-\frac{1}{2}(\frac{y^2}{\sigma_y^2} + 2yp_y\frac{\mu_{yp_y}}{\sigma_y\sigma_{p_y}}+\frac{p_y^2}{\sigma_{p_y}^2}))} 
\end{eqnarray}
The parameters in the above distribution are chosen to be RMS matched through
the FODO lattice including the space-charge effects.
\begin{figure}[!htb]
   \centering
   \includegraphics*[angle=0,width=80mm,height=57mm]{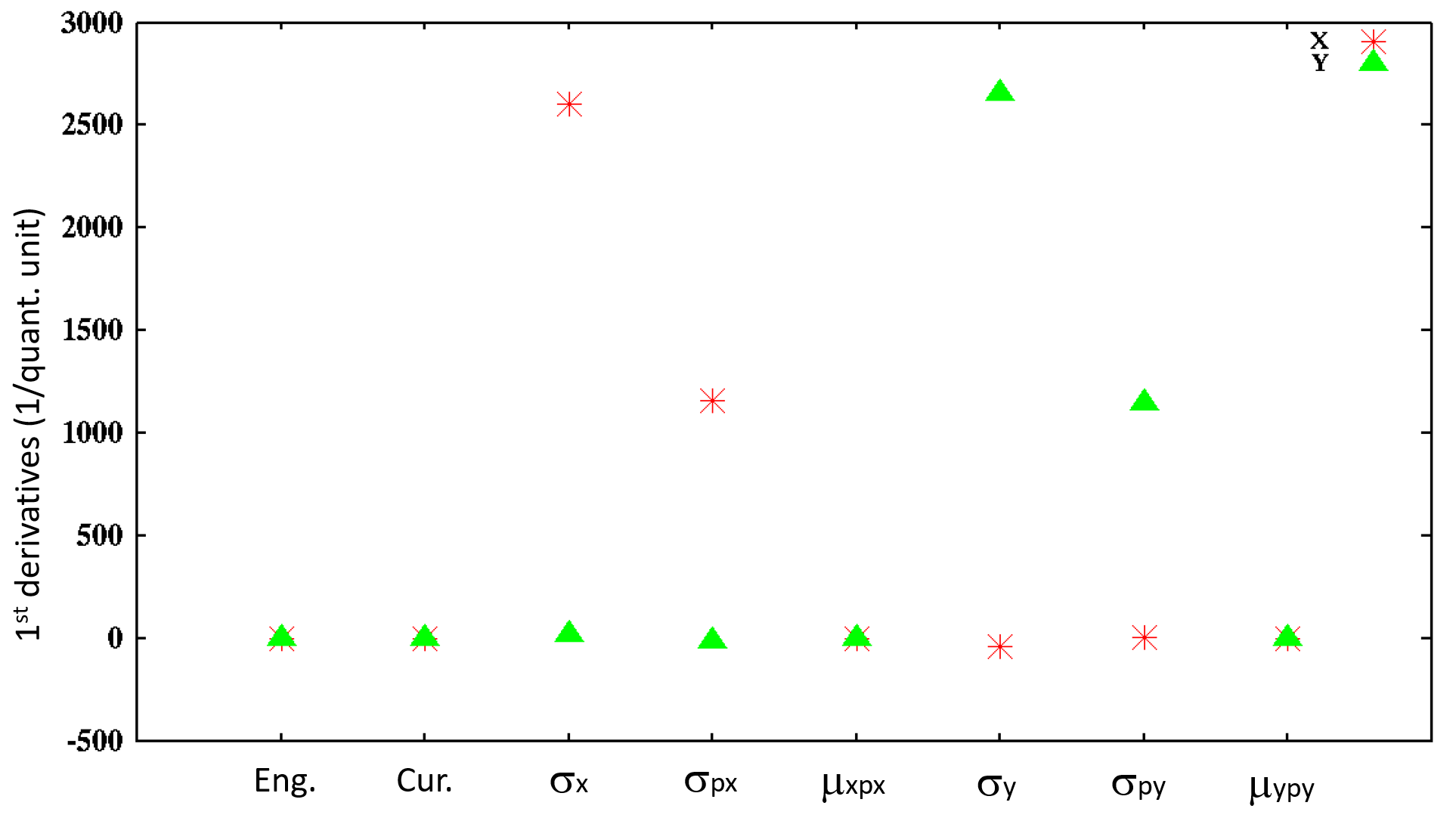}
   \caption{
Derivatives of final horizontal and vertical emittance
	with respect to eight beam parameters.}
   \label{fig3}
\end{figure}

We first checked the sensitivity of the final emittances with respect to
seven FODO lattice parameters, i.e. lengths of the drift and quadrupole elements,
and focusing strengths of two quadrupole elements.
In the this example, we used $5000$ macroparticles and $12\times 12$
spectral modes in the space-charge simulation.
The sensitivity is measured by the first derivatives of the final
emittances (normalized by initial emittance) with respect to these lattice parameters.
Figure~2 shows the sensitivities of the final horizontal and vertical emittances
with respect to the seven lattice parameters from the single
differentiable self-consistent space-charge simulation and from
the numerical finite difference approximation to the first derivative 
using eight space-charge simulations.
Here, the seven lattice parameters are drift one length (D1 L),
quadrupole one length (Q1 L), quadrupole one focusing strength (Q1 k),
drift two length (D2 L), quadrupole two length (Q2 L), quadrupole two
strength (Q2 k), drift three length (D3 L).
It is seen that the first derivatives calculated from the differentiable
space-charge model and from the finite difference approximation agree
with each other quite well. This provides a verification of the
differentiable self-consistent space-charge simulation model.
From this figure, one can see that the final emittance is much more 
sensitive to the length of the first quadrupole than the other lattice
parameters such as drift lengths and quadrupole strengths.
The change of the first quadrupole length
causes significant beam envelope variation and 
results in large final emittance change. 

Next, we checked the sensitivity of the final proton beam emittance 
with respect to the initial beam parameters.
These beam parameters are proton beam energy (Eng.), beam
current (Cur.), and six 
beam distribution parameters ($\sigma_x$, $\sigma_{p_x}$,
$\mu_{xp_x}$, $\sigma_y$, $\sigma_{p_y}$,
and $\mu_{yp_y}$). 
These parameters were represented using TPSA variables.
A number of macroparticle coordinates (also in TPSA variables)
were generated from the regular sampling method using the
beam distribution parameters.
The quadrupole magnetic field gradient was used instead
of the focusing strength since the latter includes the proton beam
energy. The change of the beam energy affects both the transverse
focusing and the space-charge effects.
Figure~3 shows the first derivatives of the final emittances with 
respect to the above eight parameters.
It is seen that the final emittances are more sensitive to the
initial distribution parameter $\sigma$s. The perturbation of these parameters
affects the matching of the initial distribution to the
FODO lattice and can cause significant emittance growth with
a mismatched beam~\cite{tom,qianghalo}.

\begin{figure}[!htb]
   \centering
   \includegraphics*[angle=0,width=80mm]{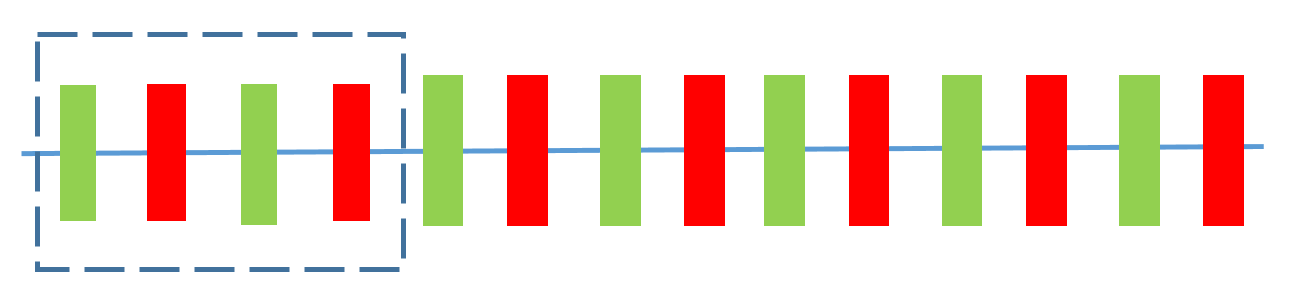}
   \caption{
Schematic plot of a FODO lattice used in the second application example.
The four quadrupoles inside the dashed line box were used to match
	the initial distribution into the periodic FODO lattice.}
   \label{fig4}
\end{figure}
The differentiable self-consistent space-charge simulation through the 
accelerator
lattice produces the final beam properties 
and their derivatives with respect to the lattice control parameters
at the exit of the accelerator
as shown in the first example.
These derivatives can be used in a gradient-based parameter optimizer 
for accelerator lattice control parameter optimization.
In the following example, we integrated the differentiable 
self-consistent space-charge simulation model
into a conjugate gradient optimizer to attain the quadrupole strengths
inside a matching section in front of a periodic FODO lattice.

A schematic plot of the matching section lattice and the periodic 
FODO lattice is shown in Fig.~4. The first four quadrupoles in 
the figure were used
to match an initial distribution to the given Twiss parameters 
at the entrance to the periodic FODO lattice. 
The Polak-Ribiere conjugate gradient optimization method~\cite{nr} was used
to minimize the objective function that is defined as follows:
\begin{eqnarray}
	f({\bf k}) & = & \frac{(\beta_x({\bf k})-\beta_{xt})^2}{\beta_{xt}^2} + 
(\alpha_x({\bf k})-\alpha_{xt})^2 + \frac{(\beta_y({\bf k})-\beta_{yt})^2}{\beta_{yt}^2} +
(\alpha_y({\bf k})-\alpha_{yt})^2 
\end{eqnarray}
where ${\bf k}$ is a set of control variables,
$\alpha_{xt}$, $\beta_{xt}$, $\alpha_{yt}$, and $\beta_{yt}$ are
the target Twiss parameters at the entrance to the periodic lattice, and
the $\alpha_{x}$, $\beta_{x}$, $\alpha_{y}$, and $\beta_{y}$ are the
beam Twiss parameters calculated from the self-consistent space-charge
simulation. These calculated beam Twiss parameters depending on the 
focusing strengths
of the quadrupoles inside the matching section. These strengths
are control variables in the above objective function.
Using the above differentiable space-charge simulation model,
the first derivatives of the objective function with respect to the
four control variables were obtained in addition to the objective
function value. These derivatives were used to construct a
conjugate direction of the gradient direction to guide the search for the
minimum solution.
\begin{figure}[!htb]
    \centering
    \includegraphics*[angle=0,width=70mm,height=50mm]{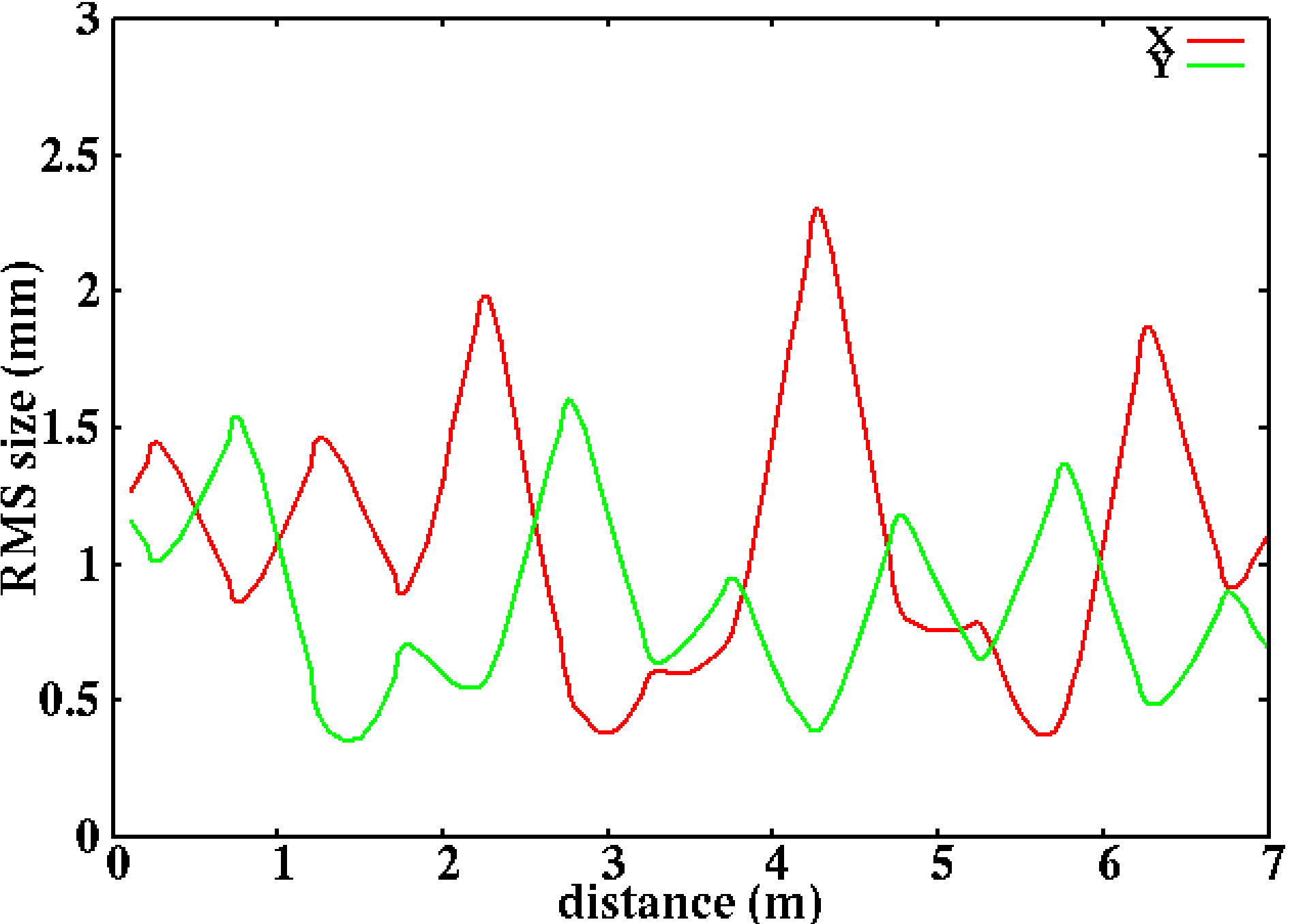}
    \includegraphics*[angle=0,width=74mm,height=51mm]{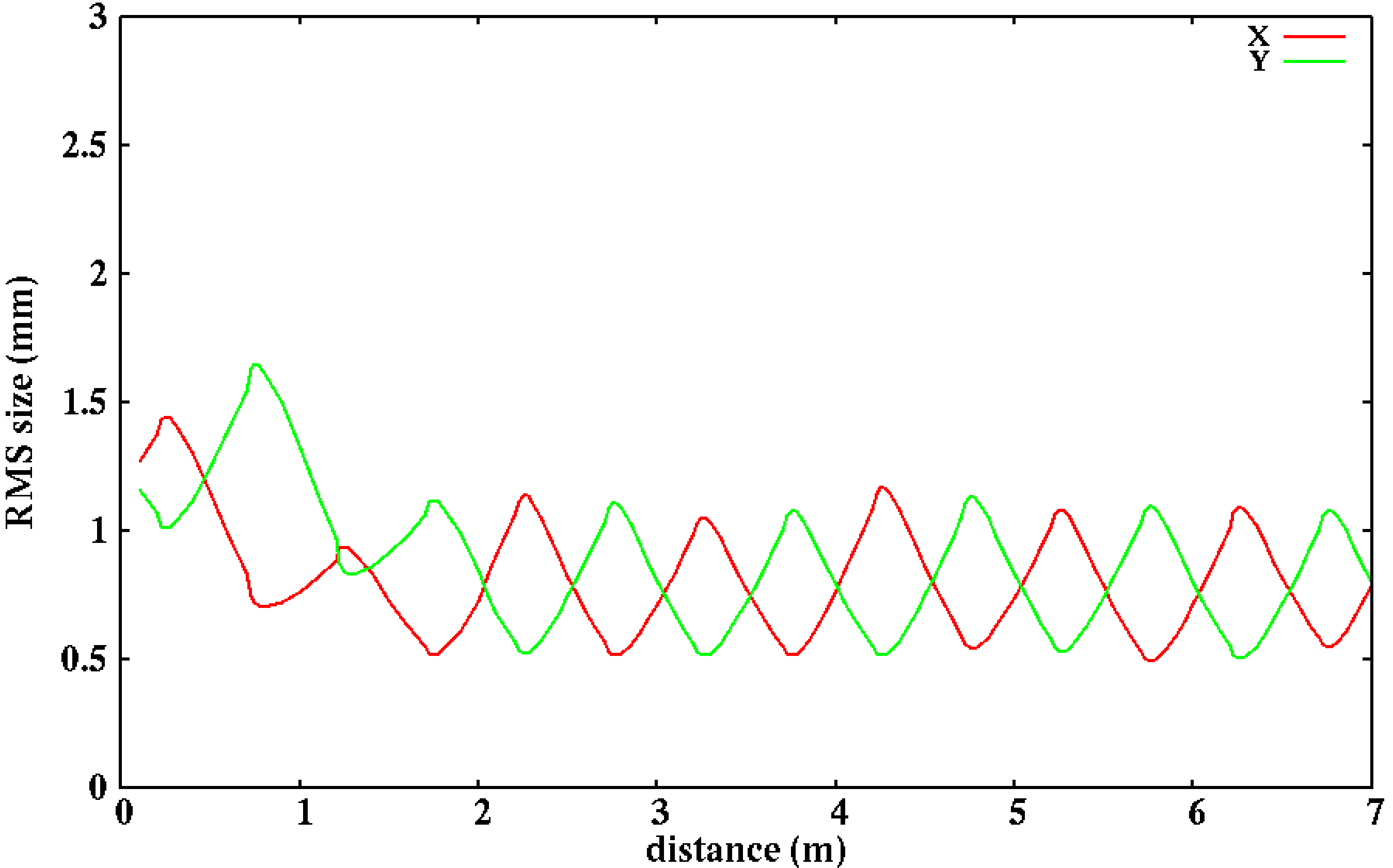}
    \caption{Transverse RMS size evolution without the quadrupole
	matching (left) and with the quadrupole matching including
	the space-charge effects (right) through the FODO lattice.}
    \label{match}
\end{figure}

Figure~\ref{match} shows the proton beam transverse RMS size evolution
through the above FODO lattice without the quadrupole matching
and with the quadrupole matching including the space-charge effects.
Here the quadrupole strengths inside the matching section
without the matching were set based
on the zero current matched solution. 
It is seen that with a $200$ Ampere beam, the space-charge effects
are significant so that the initial zero-current matched quadrupole strengths
no longer produce a matched RMS evolution inside the periodic lattice.
After reoptimizing the four quadrupole strengths including the 
space-charge effects through the self-consistent simulations,
the RMS evolution inside the periodic lattice becomes well-matched
and results in much less emittance growth (less than $10\%$)
than that from the mismatched quadrupole setting (greater than $50\%$).

\section{Conclusions}

The self-consistent space-charge simulation is an important
part in the high intensity, high brightness accelerator design.
In this paper, we proposed a differentiable self-consistent
space-charge model that can be used to efficiently study
the sensitivity of the final beam properties with respect to
the accelerator design parameters.
Using the differentiable self-consistent space-charge model, 
only one simulation is needed to attain the derivatives of the
final beam properties with respect to all accelerator design parameters
instead of multiple simulations of the conventional space-charge model.
The resultant first derivatives measure the sensitivity of the
final beam properties with respect to those design parameters. 
Some highly sensitive machine parameters can be quickly identified after
one differentiable self-consistent space-charge simulation.

As an illustration, we presented two application examples. One
example computed the sensitivities of the final proton beam emittances
with respect to seven lattice parameters or eight beam parameters
through a single differentiable self-consistent simulation.
The second example showed that the derivatives of the objective function
with respect to the quadrupole strengths inside a matching section 
from the differentiable self-consistent space-charge simulation were
used in a conjugate gradient optimizer to attain the space-charge
matched solution to a periodic FODO lattice.
The success of these two examples shows that the differentiable
self-consistent space-charge simulation model can be a useful
tool in the accelerator design.

In this study, we used a spectral solver as an illustration
of a differentiable space-charge model. In general, some other
space-charge solvers such as fast multipole solver or Green function
solver can also be used to as the differentiable space-charge model as
long as the space-charge fields from these solvers 
are represented using TPSA variables.
Furthermore, this can be generalized beyond the space-charge
effects. The other collective effects such as beam-beam effects
can also be represented using TPSA variables. Together with the
particle coordinates and accelerator machine parameters that are 
represented using the TPSA variables, one can develop a general differentiable
simulation tool that includes a variety of collective effects
for accelerator design. Such a tool will be useful to study the
beam quality sensitivity to the accelerator machine parameters 
subject the collective effects through only one simulation.

In the present study, the computational speed of the
above differentiable space-charge solver is slow compared with 
the conventional space-charge solver due to the overhead associated
with the computation using TPSA variables in the available TPSA library.
From the discussion with the author of the TPSA library used
in this study,
the computational speed can be substantially improved
if only the first-order derivative is needed~\cite{hao2022}.
A number of performance optimization strategies can be employed
to improve the computational efficiency of the library.
This will be pursued in the future study by
working with the TPSA library developers.

\section*{ACKNOWLEDGEMENTS}
We would like to thank Drs. Y. Hao, Z. Liu for the TPSA library used in this study.
This research was supported by the U.S. Department of Energy under Contract No. DE-AC02-05CH11231, and used computer resources at the National Energy Research
Scientific Computing Center.

\end{document}